\documentclass[12pt]{iopart}

\usepackage{subfig}
\usepackage{graphicx}
\usepackage{cite}
\usepackage{hyperref}

\begin{document}

\title[Jet Embedding in Heavy-Ion Collisions (ALICE)]{Fast Embedding of Jets in Heavy-Ion Collisions for Background Studies with ALICE}

\author{B Bathen, for the ALICE Collaboration}

\address{Wilhelm-Klemm-Strasse 9, 48149 M\"unster, Germany}
\ead{Bastian.Bathen@cern.ch}

\begin{abstract}
Jet reconstruction in heavy-ion collisions is strongly affected by soft background from the underlying event. For an appropriate interpretation of the jet observables it is essential to understand the influence of the background and its fluctuations on the reconstructed jets. With this purpose we study random cones and the response of a known probe embedded in a heavy-ion events. The embedded probe can be a single high-$p_{\mathrm{T}}$ track or a jet from a simulated or real pp event. This allows a detailed study of background fluctuations and verification of the performance of background subtraction methods.
\end{abstract}

\section{Introduction}
We present our results on region-to-region energy fluctuations of the soft background in Pb--Pb collisions at the LHC ($\sqrt{s_{\mathrm{NN}}} = 2.76$\,TeV) measured with the ALICE experiment. The studies are based on charged tracks with a low $p_{\mathrm{T}}$ cut of $150$\,MeV/$c$. We used two different methods \it random cones \rm and \it fast embedding \rm to investigate the background. For \it fast embedding \rm we discuss the case of embedded single high-$p_{\mathrm{T}}$ tracks.

A good knowledge of the background fluctuations is mandatory for the interpretation of jet measurements. Recently, for example the impact of background fluctuations is of particular interest since the first measurements of imbalanced jets in Pb--Pb collisions at the LHC have been published~\cite{ATLASJets}\cite{CMSJets}\cite{cacciari}.

\section{Background Subtraction}
\begin{figure}[htpb]
\begin{center}
\subfloat[][]{\includegraphics[width=0.5\textwidth]{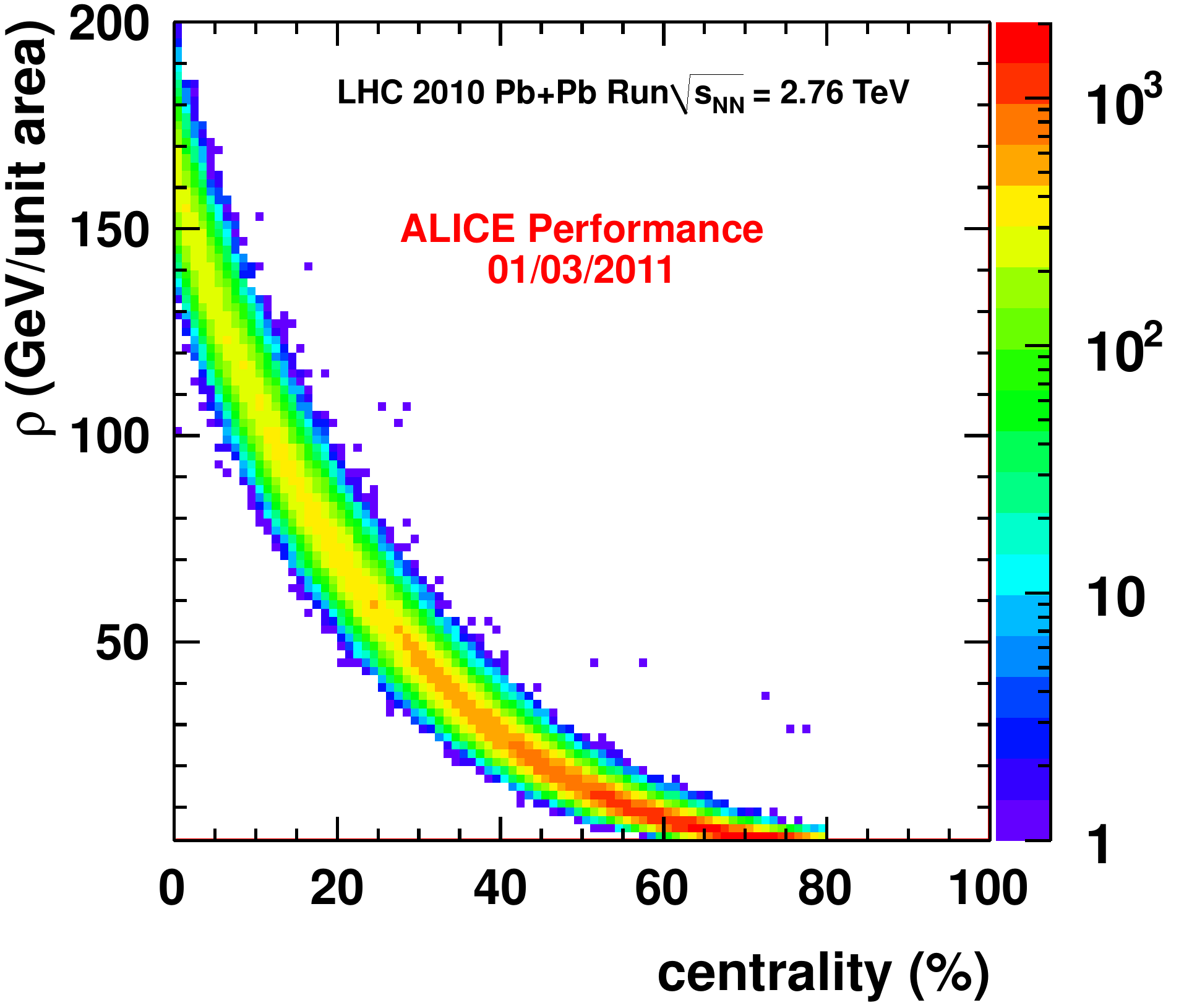}}
\subfloat[][]{\includegraphics[width=0.5\textwidth]{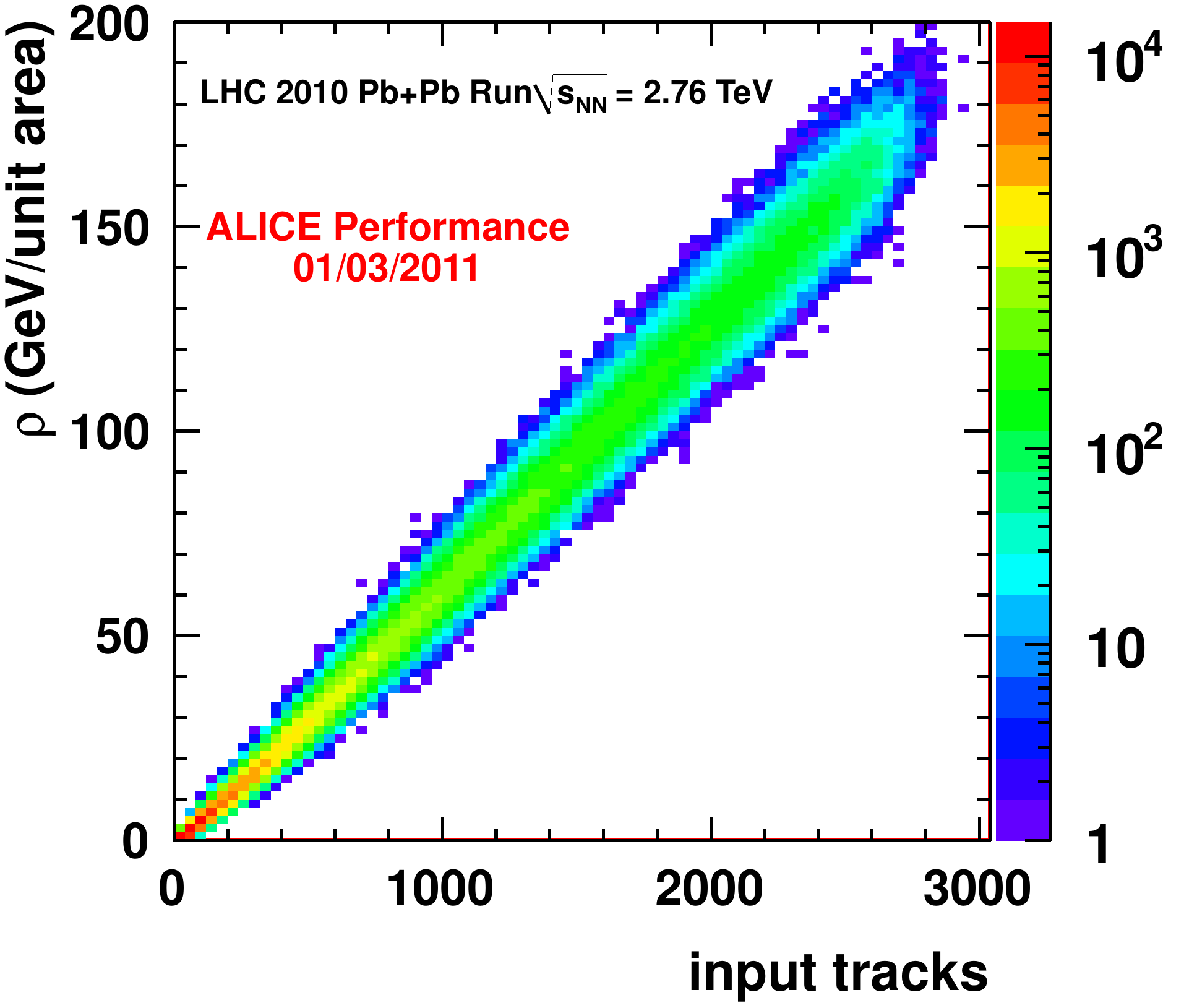}}
\end{center}
\caption{
Background density $\rho$ in Pb--Pb collisions at $\sqrt{s} = 2.76$\,TeV as function of centrality (a) and multiplicity (b) with low track $p_{\mathrm{T}}$ cut off at $150$\,MeV/$c$. Multiplicity is in this case the raw (uncorrected) number of tracks in the event, which is the number of input tracks for the jet finder.}
\label{fig:rho}
\end{figure}

The average background has to be estimated and subtracted on event-by-event basis. For this purpose, the heavy-ion event is clusterized with the $\rm{k_{T}}$-algorithm from the FastJet package \cite{fastjet}. We assume that most of these clustered objects consist of soft background, even though this is not the case for all of them. We call these \it background jets \rm in the following. The clustering procedure allows to eliminate regions with strong deviations from the average background, as we determine the background density $\rho$ of the event as the median of those background jets: $$\rho=\mathrm{median}(p_{\mathrm{T}}^{i}/A_{jet}^{i}),$$
where $p_{\mathrm{T}}^{i}$ is the momentum of the background jet $i$ with the jet area $A_{jet}^{i}$ in the $\eta-\phi$ plane. For the median we exclude the two hardest jets. In figure~\ref{fig:rho} $\rho$ is shown as function of centrality (a) and multiplicity (b). The background density increases linearly with the raw number of tracks\footnote{Raw number of reconstructed charged tracks after quality cuts, which is the number of input tracks for the jet finder.} (multiplicity).

$\rho \times A_{jet}$ is the expected amount of background energy which contributes to the jet momentum. For a jet radius of $R=0.4$ and corresponding jet area of $A_{jet}=0.5$ it is above $50$\,GeV/$c$ and up to $100$\,GeV/$c$ for the $10$\,\% most central events. The average background $\rho$ is estimated event-by-event and $\rho \times A_{jet}$ is subtracted from the reconstructed jet momentum. What still persists are the region-to-region background fluctuations $\delta p_{\mathrm{T}}$, which can not be corrected event-by-event and needs to be convoluted. The jet energy resolution is limited by these fluctuations~\cite{jetRecoHIC}:
$$p_{\mathrm{T},jet} = p_{\mathrm{T},jet}^{rec} - \rho \times A_{jet} - \delta p_{\mathrm{T}}.$$

\section{Jet Reconstruction}
The $k_{\mathrm{T}}$ algorithm~\cite{kt1}\cite{kt2} sequentially combines the tracks $i$ and $j$ with the smallest distance parameter $d_{ij}=\mathrm{min}(p_{\mathrm{T},i}^2,p_{\mathrm{T},j}^2)\frac{\Delta R_{i,j}^2}{R^2}$, where $\Delta R_{i,j}=\sqrt{(\Delta\eta)^2+(\Delta\phi)^2}$ and $R$ is the radius parameter. The $k_{\mathrm{T}}$ algorithm is especially convenient for background jets. Since the $d_{i,j}$ is weighted with the squared transverse momentum $p_{\mathrm{T},(i,j)}^2$, it basically starts clustering with low-$p_{\mathrm{T}}$ tracks and it accordingly is adaptable to soft contributions. This is disadvantageous for the reconstruction of real jets in an environment of large soft background (see e.g.~\cite{jetRecoHIC},~\cite{antikt}).

The anti-$k_{\mathrm{T}}$ algorithm on the contrary weights the distance between the tracks with the inverse transverse momentum $\mathrm{min}(1/p_{\mathrm{T},i}^2,1/p_{\mathrm{T},j}^2)$, it accordingly mainly starts clustering with high-$p_{\mathrm{T}}$ particles and is much more robust against soft background in terms of area stability and back reaction~\cite{antikt}. Hence, we use the anti-$k_{\mathrm{T}}$ algorithm for real jet reconstruction in Pb--Pb collisions. The jets are reconstructed with $R=0.4$ in a pseudorapidity range according to the jet axis of $|\eta|<0.4$.

\section{Fast Embedding}
One of the methods we use to determine the background fluctuations is the embedding of a known probe into a real heavy-ion event (see e.g.~\cite{STAR}). The probe can be single high-$p_{\mathrm{T}}$ tracks or fully simulated (e.g. PYTHIA + GEANT) or real pp jet events. Here we discuss only the results from embedded single-tracks, which are taken as jets with only one (high-$p_{\mathrm{T}}$) track. This method is called \it fast embedding\rm, since the embedding is done for already reconstructed tracks and does not take into account possible effects on reconstruction, like possible merging of track clusters with influence on the tracking efficiency and resolution. 

Since we know the transverse momentum of the embedded probe we can calculate the background fluctuations $\delta p_{\mathrm{T}}$ after background subtraction and subtraction of the momentum of the embedded probe~\cite{STAR}:
$$\delta p_{\mathrm{T}} = p_{\mathrm{T},jet}^{rec} - \rho \times A_{jet} - p_{\mathrm{T},jet}^{probe}.$$
That allows us to study in detail the influence of the soft background on the reconstructed jet observables. We also can verify the performance of the background subtraction methods we use. If the average background is correctly subtracted the mean of the $\delta p_{\mathrm{T}}$ distribution is around zero, otherwise a systematic shift in $\delta p_{\mathrm{T}}$ will occur.

A certain matching of the embedded probe and the reconstructed jet in the heavy-ion event is necessary. For embedded single tracks we just match the track with the reconstructed jet which contains the track.
As additional requirement both jets (probe and reconstructed) need to be in the jet acceptance.

The single tracks are randomly embedded over full jet $\eta-\phi$ acceptance ($|\eta|<0.4$) with a flat $p_{\mathrm{T}}$ distribution from $50$\,GeV/$c$ to $250$\,GeV/$c$. We need to keep in mind that the probes are embedded in a flat centrality distribution, that does not equate to a corresponding centrality distribution of an inclusive jet spectrum, since the cross-section of jet production in a heavy-ion event increases with the number of binary collisions. 

\section{Random Cones}
An alternative method to study the background of heavy-ion events are \it random cones\rm. Here cones with an fixed size of $R=0.4$ are randomly placed in the heavy-ion event and the $p_{\mathrm{T}}$ in the cones are summed up to \it background jets\rm. The average background is subtracted as described earlier, so we directly get the background fluctuations:
$$\delta p_{\mathrm{T}} = p_{\mathrm{T,jet}}^{rndm} - \rho \times A_{jet}.$$

\section{Results}
\begin{figure}[htpb]
\begin{center}
\subfloat[][]{\includegraphics[width=0.50\textwidth]{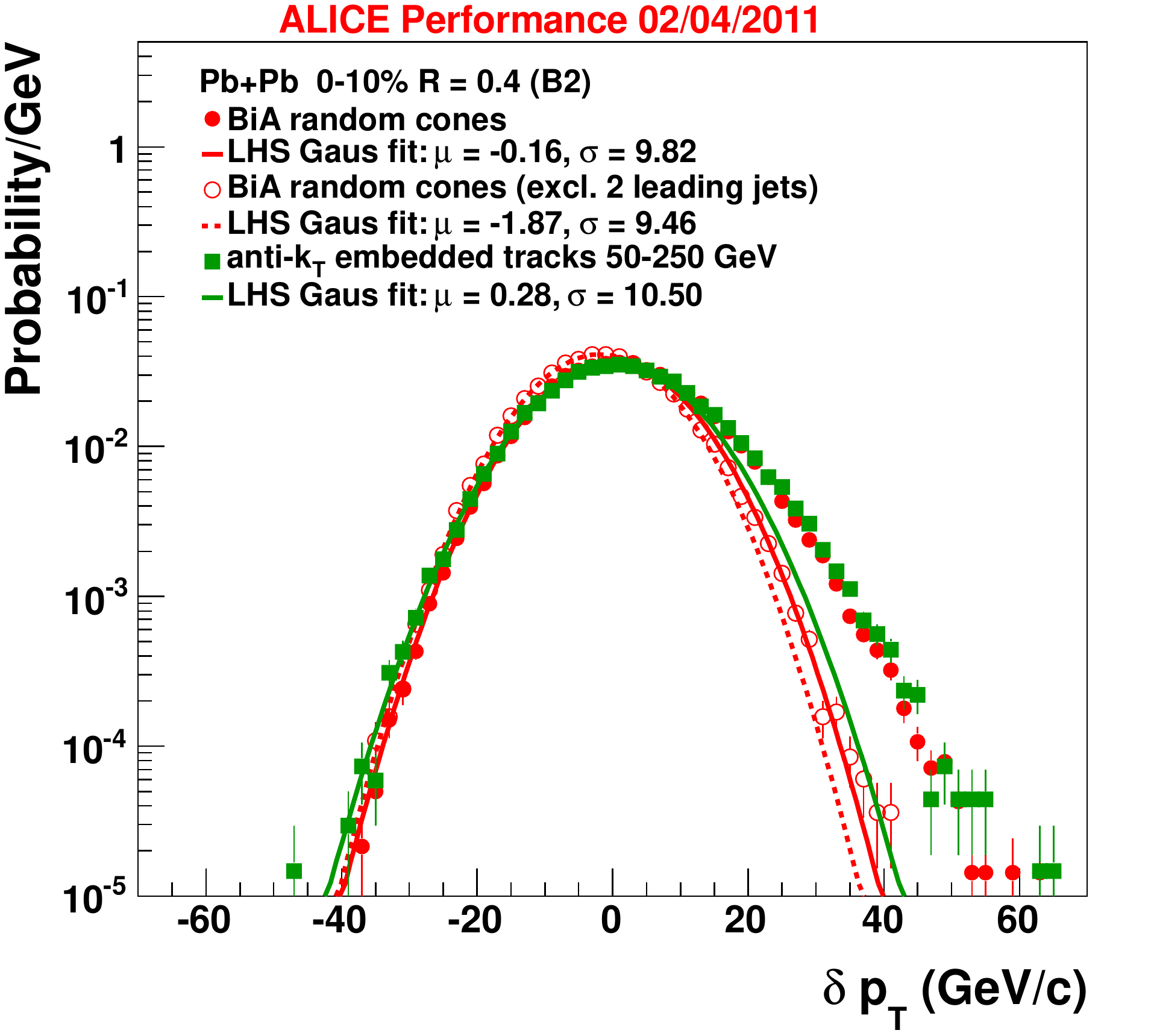}}
\subfloat[][]{\includegraphics[width=0.50\textwidth]{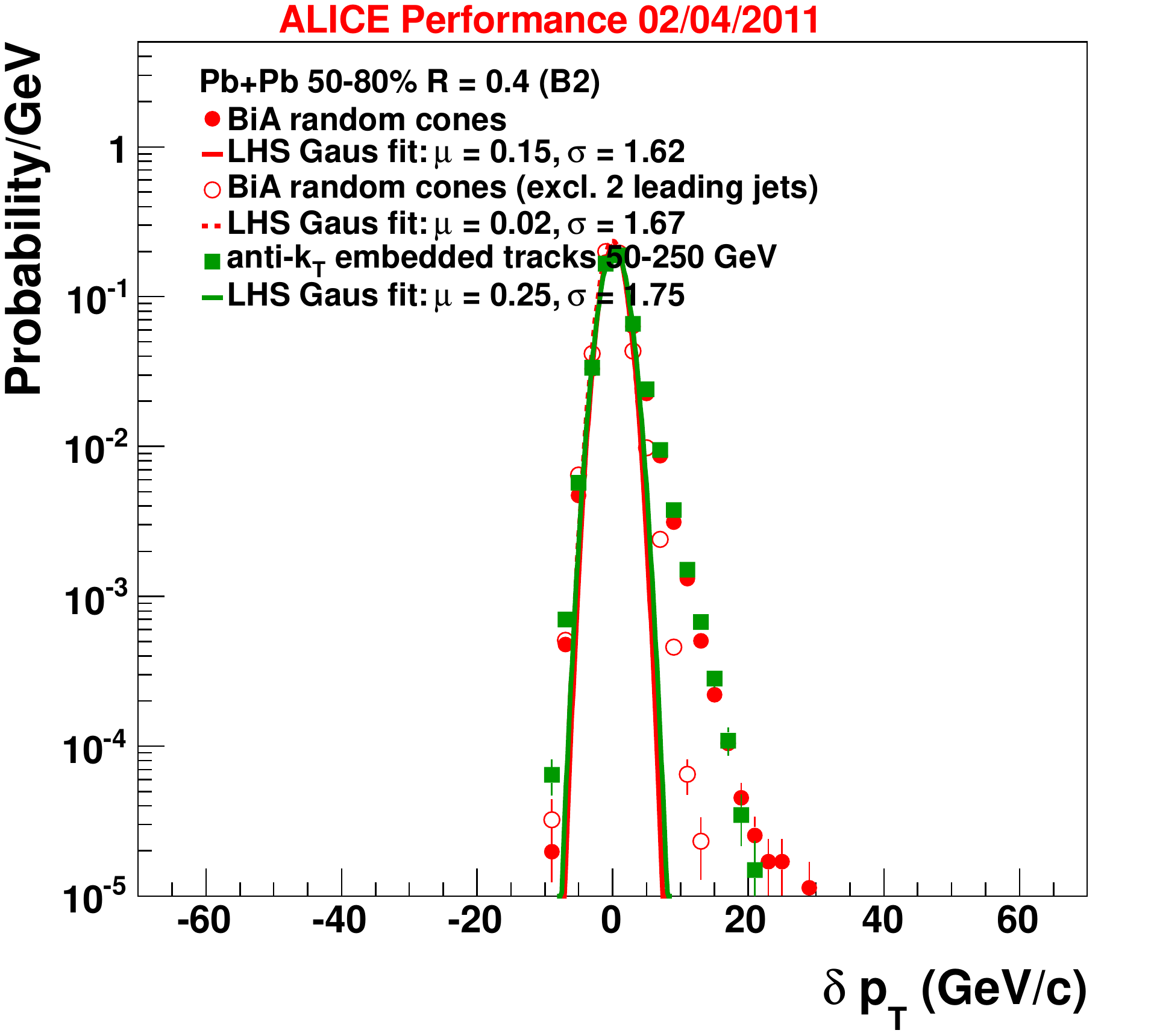}}
\end{center}
\caption{
Background fluctuations $\delta p_{\mathrm{T}}$ in Pb--Pb collisions at $\sqrt{s} = 2.76$\,TeV for central ($0-10$\,\%) (a) and peripheral events ($50-80$\,\%) (b)  with track $p_{\mathrm{T}} > 150$\,MeV/$c$.}
\label{fig:deltaPt}
\end{figure}

Figure~\ref{fig:deltaPt} shows the background fluctuations $\delta p_{\mathrm{T}}$ for two centrality classes comparison the two discussed methods \it random cones \rm and \it fast embedding \rm of single tracks. Figure~\ref{fig:deltaPt}(a) shows the results for the $10$\,\% most central events. Since the $\delta p_{\mathrm{T}}$ distribution is mainly caused by uncorrelated Poissonian fluctuations it is supposed to be a gamma-distribution\cite{tannenbaum}. However, for a large number of input tracks, which is the case for the most central events, the distribution can be well described with a Gauss in ideal case of only uncorrelated fluctuations. But, in addition jets are already part of the heavy-ion events, and those cause the tail on the right-hand side of the $\delta p_{\mathrm{T}}$ distribution. So we specify the distribution with a Gaussian fit on the left-hand side as a first assumption. More precisely it is a iterative fit with a fit range from $\mu-3\,\sigma$ to $\mu+0.5\,\sigma$ with the mean $\mu$ and width $\sigma$ of the Gaussian fit. The termination condition is a shift of $\mu$ less than $0.1$\,GeV/$c$, the maximum number of iterations is 20.

The results from \it random cones \rm and \it track embedding \rm agree pretty well. For both the mean is around zero, that demonstrates that the background subtraction method works as desired. The results demonstrate that the impact of the average background in the event to the reconstructed jet can successfully be corrected. This also is valid for all centrality classes with different amount of background, as it is shown in figure~\ref{fig:deltaPt}(b) for peripheral events (centrality $50-80$\,\%).

The width of $\delta p_{\mathrm{T}}$ is about $10$\,GeV/$c$ in the most central events ($0-10$\,\%) and decreases for more pheripheral events ($50-80$\,\%), where it is about $1-2$\,GeV/$c$ only. However, the Gaussian fit does not describe the distribution very well in such peripheral events with low number of tracks.

In addition to the random cones from all (background) jets the $\delta p_{\mathrm{T}}$ distribution from the random cones without the two leading jets is shown (fig.\ref{fig:deltaPt}, open circles). Here, the tail on the right-hand side almost disappears. This indicates that the tail indeed comes from some jets in the heavy-ion events and are not part of the general background contribution.

\begin{figure}[htpb]
\begin{center}
\subfloat[][]{\includegraphics[width=0.50\textwidth]{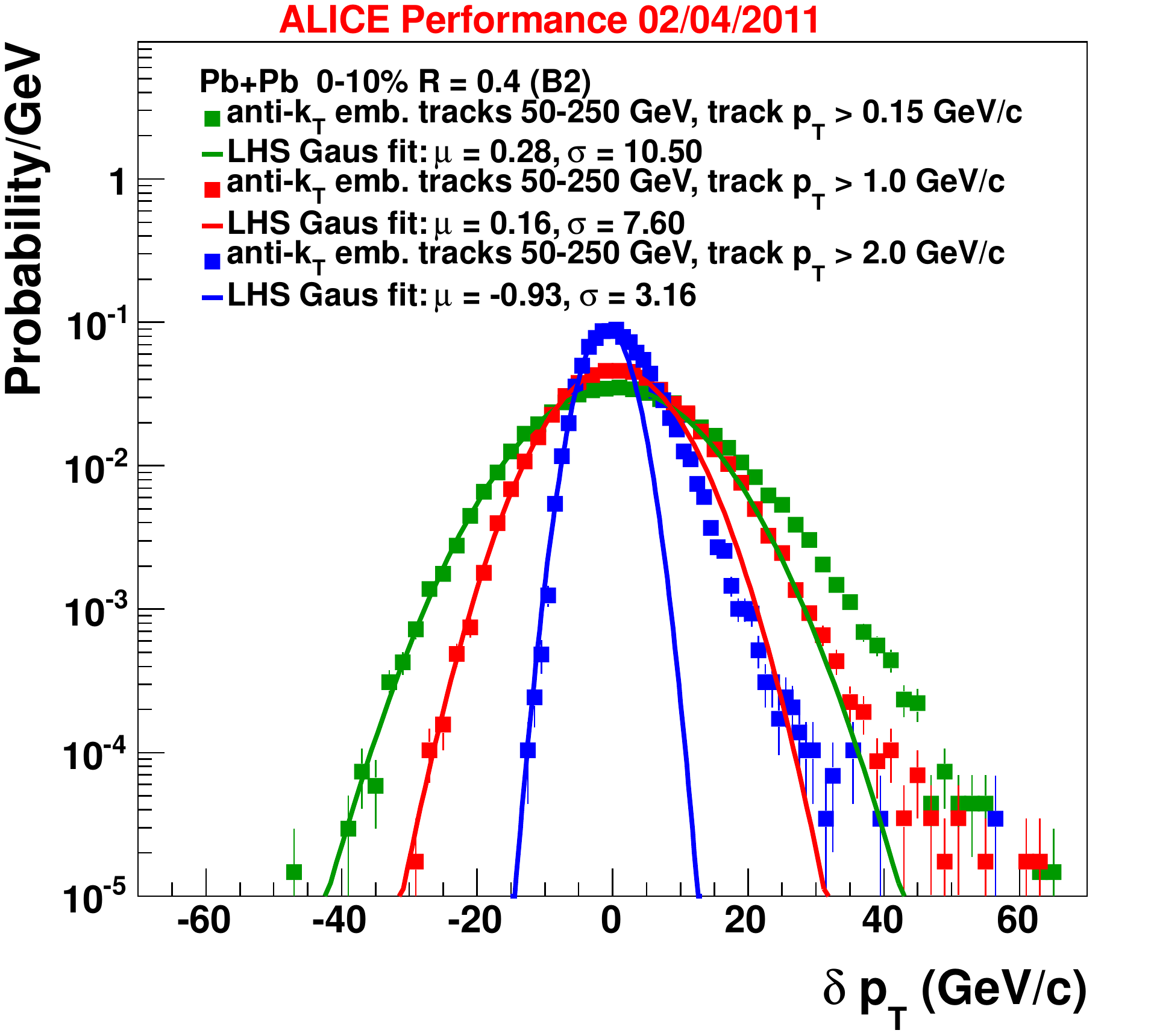}}
\end{center}
\caption{
Background fluctuations $\delta p_{\mathrm{T}}$ in central collisions ($0-10$\,\%) for track $p_{\mathrm{T}}>0.15$\,GeV/$c$, $p_{\mathrm{T}}>1.0$\,GeV/$c$ and $p_{\mathrm{T}}>2.0$\,GeV/$c$.}
\label{fig:deltaPtptcuts}
\end{figure}

The measured background, of course, directly depends on the chosen low $p_{\mathrm{T}}$ cut off. In the studies presented so far we benefit from the good tracking capabilities of ALICE for tracks of very low $p_{\mathrm{T}}$ down to $150$\,MeV/$c$. However, in view of jet reconstruction a higher $p_{\mathrm{T}}$ cut can help to reduce the influence of the background. For the track-$p_{\mathrm{T}}$ cut of $0.15$\,GeV/$c$, which we used so far, is in centrality class $0-10$\,\% the average $\rho=136$\,GeV/$c$, it decreases to $\rho=61$\,GeV/$c$ for $p_{\mathrm{T}}>1.0$\,GeV/$c$ and further to $\rho=13$\,GeV/$c$ for $p_{\mathrm{T}}>2.0$\,GeV/$c$. But the main advantage is the smaller background fluctuations which we can expect due to the reduced number of tracks. The effect for those track-$p_{\mathrm{T}}$ cuts is shown in figure~\ref{fig:deltaPtptcuts}. For $p_{\mathrm{T}}>2.0$\,GeV/$c$ is $\sigma=3.2$\,GeV, while the mean of the distribution basically stays around zero. That means that the background subtraction method also works for higher $p_{\mathrm{T}}$ cuts. Altough, we enforce the bias on a hard fragmentation, and make the situation worse for the reconstruction of quenched jets accordingly, if we raise the $p_{\mathrm{T}}$ cut. That we have to consider.

\section{Conclusion}
We presented the first measurement of the background fluctuations in heavy-ion events for charged tracks with a low $p_{\mathrm{T}}$ cut-off of $150$\,MeV/$c$. Based on the different methods \it random cones \rm and \it fast embedding \rm we see a good agreement with a background fluctuation of $\sigma\approx10$\,GeV/$c$ in the most central events. The subtraction of the average background is well under control also for events with large number of tracks. To reduce the fluctuations for the reconstruction of jets in a lower $p_{\mathrm{T}}$ region it might be necessary to introduce a higher track-$p_{\mathrm{T}}$ cut.

\section{Outlook}
We did not discuss our studies of the background fluctuations as function of multiplicity. The dependence of the number of tracks allows a comparison with the Poissonian limit. The background fluctuations are expected to approach with the Poissonian limit in case they are caused by uncorrelated sources only. 

But, also other effects contribute to a broadening of the fluctuations. E.g. we estimate the background fluctuations for different orientations to the event plane for this purpose.

Furthermore we investigate the background fluctuations with embedding of full PYTHIA jet events and quenched jets (QPYTHIA and PYQUEN). With those probes our sample becomes more realistic, but also more biased according to the jet finder, so we observe jet splitting and have signs of back reaction.

\section*{References}
\bibliographystyle{hunsrt}
\bibliography{bathen_highpt11}

\end{document}